\newcommand{\ep}{\varepsilon}
\newcommand{\nn}{\nonumber}

\newcommand{\SCR}[1]{{\mathscr #1}}

\newcommand{\CAL}[1]{{\cal #1}}

\newcommand{\J}[1]{\left\langle #1 \right\rangle}
\newcommand{\D}[1]{{\mathscr D}( #1 )}

\documentclass[preprint,12pt]{article}

\usepackage[top=30truemm, bottom=30truemm, left=25truemm, right=25truemm]{geometry}

\usepackage{fancyhdr}
\usepackage{amsthm}
\usepackage{amsmath,amssymb,latexsym,amsfonts,mathrsfs}

 \newtheorem{thm}{Theorem}[section]
 
 \newtheorem{Ass}[thm]{Assumption}
 \newtheorem{lem}[thm]{Lemma}
 
 \newtheorem{prop}[thm]{Proposition}
 \newtheorem{rem}[thm]{Remark}
 
 \numberwithin{equation}{section}

\newcommand{\Proof}[2][Proof]{
\begin{proof}[{\bf #1}]
#2
\end{proof}
}

\usepackage{color}

\usepackage{amssymb}

\begin{document}

\begin{flushleft}
{ \Large \bf Nonexistence of wave operators via strong propagation estimates for Schr\"{o}dinger operators with sub-quadratic repulsive potentials}
\end{flushleft}

\begin{flushleft}
{\large Atsuhide ISHIDA}\\
{
Katsushika Division, Institute of Arts and Sciences, Tokyo University of Science, 6-3-1 Niijuku, Katsushika-ku,Tokyo 125-8585, Japan\\ 
Email: aishida@rs.tus.ac.jp
}
\end{flushleft}
\begin{flushleft}
{\large Masaki KAWAMOTO}\\
{Department of Engineering for Production, Graduate School of Science and Engineering, Ehime University, 3 Bunkyo-cho Matsuyama, Ehime, 790-0826. Japan }\\
Email: {kawamoto.masaki.zs@ehime-u.ac.jp}

\end{flushleft}
\begin{abstract}
 Sub-quadratic repulsive potentials accelerate quantum particles and can relax the decay rate in the $x$ of the external potentials $V$ that guarantee the existence of the quantum wave operators. In the case where the sub-quadratic potential is $- |x|^{\alpha} $ with $0< \alpha < 2$ and the external potential satisfies $|V(x) | \leq C (1+|x|) ^{-(1- \alpha /2) - \ep} $ with $\ep>0$, Bony {\it et al}. [J. Math. Pures Appl., \textbf{84}, 509 (2005)] determined the existence and completeness of the wave operators, and Itakura [J. Math. Phys., \textbf{62}, 061504 (2021)] then obtained their results using stationary scattering theory for more generalized external potentials. Based on their results, we naturally expect the following. If the decay power of the external potential $V$ is less than ${ -(1- \alpha /2) } $, V is included in the short-range class. If the decay power is greater than or equal to ${ -(1- \alpha /2) } $, $V$ is included in the long-range class. In this study, we first prove the new propagation estimates for the time propagator that can be applied to scattering theory. Second, we prove that the wave operators do not exist if the power is greater than or equal to $-(1- \alpha /2)$ and that the threshold expectation of ${ -(1- \alpha /2) } $ is true using the new propagation estimates.
\end{abstract}

\begin{flushleft}


\end{flushleft}

\section{Introduction}

Consider the free Hamiltonian as a self-adjoint operator acting on $L^2({\bf R}^n)$:
\begin{align*}
H_0 = p^2 - \sigma |x|^{\alpha}, 
\end{align*}
where $x= (x_1, x_2,...,x_n) \in {\bf R}^n$, $p = -i \nabla $, $p^2 = p \cdot p = - \Delta $, $\sigma > 0$ and $0 < \alpha < 2$.  
The external potential $V$ is defined as follows. 
\begin{Ass}\label{A1}
Let $V$ be a multiplication operator of the function $V\in C^{\infty} ({\bf R}^n)$ that satisfies the following decaying conditions: for $0 < \theta \leq  \rho := 1- \alpha /2$, $|x| \gg 1$ and any multi-index $\beta$, there exist constants $C_{V,\beta} >0$ such that 
\begin{align*}
\left| \partial ^{\beta} V (x)  \right| \leq C_{V,\beta} \J{x}^{- \theta  - |\beta|} , 
\end{align*}
where $\J{x} := (1 + |x|^2) ^{1/2}$. Moreover, there exist $0 < c_0 < C_0$ such that
\begin{align*}
 c_0 \J{x}^{-\theta}  \leq V(x) \leq C_0 \J{x}^{- \theta }
\end{align*}
 or 
\begin{align*}
 c_0 \J{x}^{-\theta }  \leq - V(x) \leq C_0 \J{x}^{-  \theta  }
\end{align*} 
holds for $|x| \gg 1$.
\end{Ass}
Under this assumption, we define the perturbed Hamiltonian $H = H_0 +V$, which is also a self-adjoint because $V$ is bounded. Then, a family of unitary operators can be defined as follows: 
\begin{align*}
W(t) := e^{itH} e^{-itH_0}, \quad t \in {\bf R},
\end{align*}
owing to the self-adjointness of $H_0$ and $H$. If the external potential $V\in L^{\infty} ({\bf R}^n)$ satisfies
\begin{align}\label{0} 
\left| 
V(x)
\right| \leq C \J{x}^{- \rho - \ep}
\end{align}
with $\ep>0$, then the existence and completeness of wave operators 
\begin{align*}
{W}^{\pm} := \mathrm{s-} \lim_{t \to \pm \infty} W(t) 
\end{align*}
can be proven (see Bony-Carles-H\"{a}fner-Michel \cite{BCHM} and Itakura \cite{It2}). Hence, the external potential satisfying \eqref{0} can be considered short-range. Physically, the repulsive potentials $- \sigma |x|^{\alpha} $ accelerate the quantum particle, and the probability of the position $x(t)$ and velocity $v(t)$ of the particle behave similar to $\CAL{O} (t^{1/ \rho})$ and $\CAL{O}(t^{1/\rho -1})$, respectively (see Section 1.3 of Itakura \cite{It3}). This acceleration phenomenon changes the threshold of the decay power of the external potential for which the wave operators exist. In the case where $\sigma = 0$, Dollard \cite{Do}, Jensen-Ozawa \cite{JO}, and others have determined that the threshold for the existence of the wave operators is $\rho =1$ and that the wave operators do not exist if $\rho \leq 1$. Subsequently, Ozawa \cite{O} considered the case a Stark Hamiltonian,  which is closely related to a case where $\sigma \neq 0$ and $\alpha =1$, and showed that its threshold is $\rho =1/2$. Ishida \cite{Is} considered the case where $\alpha =2$, showed that its threshold cannot be characterized by the polynomial decay of the external potential, and determined that $(\log (1+|x|))^{-1}$ is the threshold of the decay rate. Based on such studies, the threshold in our case is reasonably $\rho = 1- \alpha /2$.  Hence, we prove that this expectation is true using the following theorem.
\begin{thm}\label{T1}
Under Assumption \ref{A1}, the wave operators ${W}^{\pm}$ do not exist.
\end{thm}

The key estimate to demonstrate this theorem is the {\em strong propagation estimate} for $e^{-it H}$, which plays an important role in scattering theory. A well-known approach to obtain this estimate employs the conjugate operator $\SCR{A}$ such that the commutator on $\D{H} \cap \D{\SCR{A}}$ satisfies the Mourre inequality $ \varphi(H)  i [H , \SCR{A}] \varphi(H) \geq c_0 \varphi (H) ^2$ for $\varphi \in C_0^{\infty}({\bf R})$ with a positive constant $c_0 >0$. In this study, we employ $\SCR{A}$ as follows: 
\begin{align} \label{ad10}
\SCR{A} := \J{x}^{- \alpha} x \cdot p + p \cdot x \J{x}^{- \alpha}, 
\end{align}
and this operator is different from the conjugate operators used in \cite{BCHM} and \cite{It}. Using conjugate operator $\SCR{A}$, we obtain the following theorem.
\begin{thm}\label{T2}
Let $\alpha _0 = \min \{ \alpha \sigma , (2- \alpha) \sigma \} $, $0< \delta \ll \alpha _0$, and $g \in C^{\infty} ({\bf R})$ be a cut-off function such that $g(x) = 1$ if $x < \delta $ and $g(x) = 0$ if $x > 2 \delta$. Then, for any $\kappa  \geq 0$, $\varphi \in C_0^{\infty}({\bf R})$ and $\psi \in L^2( {\bf R}^n) $, there exist $C_{\kappa} >0$ such that 
\begin{align}\label{adad30}
\left\| g(\SCR{A} /t) e^{-itH} \varphi (H) \J{\SCR{A}}^{- \kappa} \psi \right\| \leq C_{\kappa} |t|^{- \kappa} \| \psi \|
\end{align}
holds for $|t| \geq 1$. 
\end{thm}

\begin{rem}
The nonexistence of the embedded eigenvalues for $H$ has been proven by \cite{It} under weaker assumptions than Assumption \ref{A1}; hence, we have $\sigma (H) = \sigma _{\mathrm{ac}} (H) = {\bf R}$. 
\end{rem}
\begin{rem}\label{R1}
For the strong propagation estimates, Skibsted \cite{Sk} and Adachi \cite{A2} showed \eqref{adad30} for generalized frameworks with a suitable Hilbert space and a pair of self-adjoint operators ${H}$ and ${\SCR{A}}$. However, these studies had to assume that ${H}$ is bounded from below or that $i[H ,\SCR{A}]$ can be extended to a bounded operator. By contrast, for our model, $H$ and $\SCR{A}$ do not satisfy the both conditions mentioned above, and we do not rely on the results of \cite{Sk} and \cite{A2}. Hence, our estimate \eqref{adad30} is new and not a consequence of those results.
\end{rem}
\begin{rem}
In \cite{BCHM}, the authors considered  $\tilde{H} = p^2 - \sigma \J{x}^{\alpha} +V $ instead of $H = p^2 - \sigma |x|^{\alpha} +V$ because ${H}$ can be written as $H = \tilde{H} + \sigma (|x|^{\alpha} - \J{x}^{\alpha} ) $ and $|x|^{\alpha} - \J{x}^{\alpha}$ belongs to the short range. Hence, if the asymptotic completeness (or nonexistence of eigenvalues) is proven for $e^{it \tilde{H}} e^{-it(\tilde{H} -V)}$ (or $\tilde{H}$), the same conclusion is true for $H$. By using this fact, \cite{BCHM} considered only for Hamiltonian $\tilde{H}$. In this paper, we need to calculate the commutator between $H$ and its conjugate operator many times and then $ |x|^{\alpha} - \J{x}^{\alpha}$ makes calculations much difficult when we take conjugate operator as one in \cite{BCHM} (see, \S{4}). From this reason, we need to establish the suitable Morre estimates which can deal with $H$ directly without passing through $\tilde{H}$. 

\end{rem}
By Theorem \ref{T1}, we determine that the repulsive potential $- \sigma |x|^{\alpha}$ relaxes the decay rate of the external potential $V$, which guarantees the existence of the wave operators. Conversely, the deceleration phenomenon was recently found by Ishida-Kawamoto \cite{IK,IK2} when the harmonic potential $\sigma (t) |x|^2 $ exists with a time-decaying coefficient $\sigma (t)$; in this case, it was shown that there exists $\rho >1$ in \eqref{0} such that the wave operators do not exist. Regarding these studies, to consider the sub-quadratic repulsive potential with time-decaying coefficients $- \sigma (t) |x| ^{\alpha}$ seems interesting. Our results are fundamental for considering such studies. \par
In the usual method to prove the nonexistence of wave operators (e.g., \cite{Do, JO, O}), the estimate  
\begin{align}\label{ad30}
\int_{a}^b \left( V e^{-itH_0} \phi , e^{-itH_0} \phi \right) dt \geq C \int_{a}^b \frac{dt}{t}, \quad a>b \gg 1
\end{align} 
is necessary for $\phi \in C_0^{\infty} ({\bf R}^n)$. If $\sigma =0$, the well-known MDFM-type decomposition $e^{-itH_0} = \CAL{M}(t) \CAL{D}(t) \SCR{F} \CAL{M}(t)$ holds, where $\CAL{M}(t)\phi(x)=e^{ix^2/(4t)}\phi(x)$, $\CAL{D}(t)\phi(x)=(2it)^{-n/2}\phi(x/(2t))$ and $\SCR{F}$ is the standard Fourier transform of $L^2({\bf R}^d)$. Then, the problem of the nonexistence of the wave operators is whether 
$$ \mathrm{s-} \lim_{t \to \pm \infty} e^{itH} \CAL{M}(t) \CAL{D}(t) \SCR{F} \phi $$ 
do not exist. With this reduction, \eqref{ad30} can be reduced to 
\begin{align*}
\int_{a}^b \left( V \CAL{M}(t) \CAL{D}(t) \SCR{F}  \phi , \CAL{M}(t) \CAL{D}(t) \SCR{F}  \phi \right) dt \geq C \int_{a}^b \frac{dt}{t}
\end{align*}
(with some error terms). Let $\phi $ be $ \chi_{(a' \leq 2|x| \leq b')} [\SCR{F}\phi ](x)$ with a characteristic function $\chi$. We can then easily obtain 
\begin{align*}
\int_{a}^b \left( V \CAL{M}(t) \CAL{D}(t) \SCR{F}  \phi , \CAL{M}(t) \CAL{D}(t) \SCR{F}  \phi \right) dt &=
\int_{a}^b \left( V \chi _{(a't \leq |x| \leq b't)} \CAL{M}(t) \CAL{D}(t) \SCR{F}  \phi , \CAL{M}(t) \CAL{D}(t) \SCR{F}  \phi \right) dt 
\\ & \geq C \int_{a}^b \frac{dt}{t}.
\end{align*} 
These arguments are based on the integral kernel of $e^{it \Delta}$ having an explicit expression. In the case of $\sigma \neq 0$, the imitation of such an argument is difficult. Indeed, an MDFM-type decomposition for $e^{-itH_0}$ with $0< \alpha <2$ has not yet been obtained (if $\alpha=2$, the Mehler formula is known as a correspondence, e.g., \cite{Is, N}). Therefore, an alternative approach must be established. Our plans to obtain \eqref{ad30} are as follows. We first present the large-velocity estimate in Section 3. A similar estimate was proved by \cite{BCHM}. However, to show the nonexistence of wave operators, the estimate in \cite{BCHM} is insufficient, and we must extend this estimate. In particular, we show 
\begin{align}\label{ad31}
\int_1^{\infty} \left\| F \left( \frac{|x|^{\rho} }{t} \right) e^{-itH} \varphi(H) \J{x}^{- \rho} \phi \right\| ^2 \frac{dt}{t} \leq C \| \phi \|^2.
\end{align}
in Proposition \ref{P2} for a large-velocity cut-off $F$, which is {\em not} compactly supported. To employ this cut-off, we must prove auxiliary lemmas (Lemmas \ref{L1} and \ref{L2}). Next, we provide the proof of Theorem \ref{T2} in Section 5 using Proposition \ref{P2} and the Mourre inequality (Proposition \ref{P3}). As mentioned in Remark \ref{R1}, we cannot rely on the results of \cite{Sk} and \cite{A2} because operator $H$ is not bounded from below, and the commutator of $H$ and conjugate operator $\SCR{A}$ cannot be extended to the bounded operator. For these reasons, we provide some modifications of the approach used in \cite{Sk} to suite our case; the boundedness from below of $H$ are only used to show property of the domain invariance,
\begin{align}\label{adadad1}
 e^{-itH} \varphi(H) \D{\SCR{A}^N} \subset \D{\SCR{A} ^N},
\end{align}
for $N \in {\bf N}$. Hence, we must only provide a different proof for \eqref{adadad1} without using the lower boundedness of $H$. The approach is relatively simple; we simply divide $e^{-itH} \varphi(H)$ into a $ \cos (t H) \varphi(H) $ and $\sin (tH) \varphi(H)$ part, for which we can justify the Helffer-Sj\"{o}strand formula and employ a direct calculation (see \S{5}).

To complete the proof of Theorem \ref{T1}, we want to follow the approaches in \cite{Is, IK, IK2, IW, JO, O} using the strong small-velocity estimate for the free-time evolution $e^{-itH_0}$ that has the following shape:
\begin{align}\label{ad32}
\left\| 
 g\left(  \frac{|x|^{\rho}}{t} \right) e^{-itH_0} \phi
\right\| \leq C t^{-N} 
\end{align} 
for some $\phi \in \SCR{S}({\bf R}^n)$, where $g$ is the small-velocity cut-off $g$ and $N\in{\bf N}$. However, in our case, showing \eqref{ad32} is difficult even for $H_0$ by direct calculation because $e^{-itH_0}$ does not have the MDFM-type decomposition. Hence we alternatively employ Theorem \ref{T2} for $H_0$ (that is, $V\equiv0$), and then estimates $\| g(\SCR{A}/t) e^{-itH_0} \phi \| = o (t^{-1})$ and $ \| V(x) (1-g(\SCR{A}/t) ) \varphi(H_0)  \| \sim \CAL{O}(t^{-1})$ as $t \to \infty$ enable us to show Theorem \ref{T1} for $\theta=\rho$ (see \S{6}). We finally prove Theorem \ref{T1} for $0<\theta<\rho$ using the result of $\theta=\rho$ and Theorem \ref{T2} for $H$. Here we emphasize that the case of $\theta < \rho$ can be shown by the different scheme for the case of $\theta = \rho$. \par

In previous studies, proofs for the nonexistence of wave operators fully use good tools, such as MDFM-type decompositions and Fourier multipliers, that a free propagator has. However, our approach employs only the large-velocity propagation estimates and strong propagation estimates of the conjugate operator. We think such strategies are new and applicable to more developed studies.

\section{Preliminaries}

In this section, we introduce important lemmas. Throughout this study, $\|  \cdot \| $ indicates the norm on $L^2({\bf R}^n)$ or operator norm on $L^2({\bf R}^n)$, and $(\cdot , \cdot )$ indicates the inner product of $L^2({\bf R}^n)$. If an operator $A$ satisfies $\| A \| \leq C$ with a constant $C$, which is independent of any parameters under consideration, then we may denote $A$ by $B_0$, and compact operators are denoted by $C_0$. 

One difficulty in this study is the issue due to the domain; showing that $ \varphi(H) \SCR{S}({\bf R}^n) \subset \D{ p^2 + |x|^{\alpha} } $ is difficult even if $\varphi \in C_0^{\infty} ({\bf R})$. This makes many arguments difficult; hence, in this section, we show some properties of domain invariance that are necessary to show Theorem \ref{T1}. 

\begin{lem}\label{La1}
For any $\varphi \in C_0^{\infty} ({\bf R})$, $z \in {\bf C} \backslash {\bf R}$ and $j \in \{ 1,2,...,n \} $, the domain invariance
\begin{align} \label{ad1}
\J{x}^{- \alpha} (z-H)^{-1} L^2({\bf R}^n) \subset \D{p^2}, \quad \J{x}^{- \alpha /2} (z-H)^{-1} L^2({\bf R}^n) \subset \D{p_j}
\end{align}
hold. In particular, 
\begin{align} \label{ad2}
\J{x}^{- \alpha} \varphi (H) L^2({\bf R}^n) \subset \D{p^2}, \quad \J{x}^{- \alpha /2} \varphi (H) L^2({\bf R}^n) \subset \D{p_j}
\end{align}
hold.
\end{lem}
\Proof{
We show only \eqref{ad1}, because \eqref{ad2} can be shown using the Helffer-Sj\"{o}strand formula and \eqref{ad1}. Owing to the similar arguments in Lemma 2.3 of \cite{BCHM}, we have that 
\begin{align*}
 \J{x}^{-\alpha} (z-H)^{-1} &= (p^2+1)^{-1}(p^2 +1) \J{x}^{-\alpha} (z-H)^{-1} \\ &= 
 (p^2 + 1) ^{-1} \J{x}^{-\alpha} (H + \sigma |x|^{ \alpha}-V) (z-H)^{-1}  + (p^2 + 1) ^{-1} [p^2, \J{x}^{- \alpha} ] (z-H)^{-1} 
 \end{align*} 
 and that 
 \begin{align*}
 & (p^2 + 1) ^{-1} [p^2, \J{x}^{- \alpha} ] (z-H)^{-1} \\ & = (p^2 + 1) ^{-1} [p^2, \J{x}^{- \alpha} ] (p^2 +1)^{-1} (H + \sigma |x|^{ \alpha}-V) (z-H)^{-1}
 \\ &= (p^2 + 1) ^{-1} [p^2, \J{x}^{- \alpha} ] (p^2 +1)^{-1} \J{x}^{\alpha} \cdot \J{x}^{- \alpha} (H + \sigma |x|^{ \alpha}-V) (z-H)^{-1}. 
 \end{align*}
Clearly, the operator 
\begin{align*}
[p^2, \J{x}^{- \alpha} ] (p^2 +1)^{-1} \J{x}^{\alpha} = \sum_{j=1}^n B_0 \J{x}^{- \alpha -2} x_j p_j (p^2 + 1)^{-1} \J{x}^{\alpha} + B_0 (p^2 + 1)^{-1} \J{x}^{\alpha}
\end{align*} 
on $\D{\J{x}^{\alpha}}$ can be extended to the bounded operator, and this implies $\J{x}^{- \alpha} (z-H)^{-1} L^2({\bf R}^n) \subset \D{p^2}$. Next, we present the second part of \eqref{ad1} using the first term of \eqref{ad1}. 
We fix $z$ and set $u_l \in \SCR{S} ({\bf R}^n) $ such that $u_l \to (z-H)^{-1} u $ and $p_j^2 \J{x}^{-\alpha} u_l \to p_j^2 \J{x}^{-\alpha} (z-H)^{-1} u $ as $l \to \infty$. For $ \phi \in \SCR{S}({\bf R}^n)$, we have 
\begin{align*}
p_j ^2 \J{x}^{-\alpha } \phi  &= [p_j^2, \J{x}^{- \alpha /2}] \J{x}^{- \alpha /2}  + \J{x}^{-\alpha /2} p_j ^2 \J{x}^{- \alpha /2}  \phi 
\\ &= \left( B_0 + 2 i \alpha  x_j \J{x}^{\alpha /2 -2} p_j \right) \J{x}^{- \alpha}  + \J{x}^{-\alpha /2} p_j ^2 \J{x}^{- \alpha /2} \phi
\end{align*}
and hence, 
\begin{align*}
\left\| p_j \J{x}^{- \alpha /2} (u_l - u_k) \right\| ^2  \to 0 , \quad \mbox{as} \quad l, k \to \infty.
\end{align*}
Because $p_j \J{x}^{- \alpha /2} $ is a closed operator, we have $p_j \J{x}^{- \alpha /2} u_l \to   p_j \J{x}^{- \alpha /2} (z-H)^{-1} u \in L^2({\bf R}^n) $.
}

\begin{lem}\label{La2}
For all $N \in {\bf N}$ and $z \in {\bf C} \backslash {\bf R}$, we have  
\begin{align} \label{ad3}
(z-H)^{-1} \D{\J{x}^{\rho N} } \subset  \D{\J{x}^{\rho N}} , \quad \varphi(H) \D{\J{x}^{\rho N} } \subset \D{\J{x}^{\rho N}}. 
\end{align}
In particular, for any fixed $t \in {\bf R}$,
\begin{align} \label{ad4}
 e^{-itH} \varphi(H) \D{\J{x}^{\rho N}}  \subset \D{ \J{x}^{\rho N}}. 
\end{align}
\end{lem}
\Proof{
First, we show the first term of \eqref{ad3} with $N=1$ and use induction. Using the Helffer-Sj\"{o}strand formula, the second term of \eqref{ad3} can be similarly shown. Let $l \in {\bf N}$ and set $\gamma \in C_0^{\infty} ({\bf R})$ such that $\gamma(t) =1$ if $|t| \leq 1$ and $\gamma(t) = 0$ if $|t| >2$ and $J_l (x) = \J{x}^{\rho} \gamma (\J{x}/l) $. Then, by Lemma \ref{La1}, we have $  J_l (x) (z-H)^{-1} L^2({\bf R}^n) \subset \D{p^2 + |x|^{\alpha}} $, which implies that the commutator $[J_l (x) , (z- H)^{-1}]$ can be calculated as 
$$ 
(z-H) ^{-1} [H, J_l (x)]  (z- H)^{-1} = 
(z-H) ^{-1}  [p^2, J_l (x)]  (z- H)^{-1}
$$ since $\overline{p^2 - \sigma |x|^{\alpha} } = \overline{p^2  } -\sigma  \overline{|x|^{\alpha} }  $ holds on $\D{p^2 + |x|^{\alpha}} $. Hence, for $u_l := J_l({x}) (z-H)^{-1} \J{x}^{- \rho} u $, $u \in L^2({\bf R}^n)$, we have 
\begin{align*}
u_l &= - (z-H)^{-1} [ H, J_l (x) ] (z-H)^{-1} \J{x}^{- \rho} u + (z-H)^{-1} \gamma (\J{x}/l ) u \\ 
& = -  (z-H)^{-1} B_0 \times \left(\sum_{m=1}^n p_m x_m \J{x}^{-\alpha /2 -1} +O(1)+ O(l^{-1- \alpha /2}) \right)
(z-H)^{-1} \J{x}^{- \rho} u  
\\ & \qquad \quad + (z-H)^{-1}\gamma (\J{x}/l ) u
\end{align*}
converges as $l \to \infty$, and this implies $ (z-H)^{-1} \D{\J{x}^{\rho } } \subset  \D{\J{x}^{\rho }}$.

Subsequently, suppose that $ (z-H)^{-1} \D{\J{x}^{\rho k} } \subset  \D{\J{x}^{\rho k}}$ for some $k \in {\bf N}$. Then, by defining $v_l = J_l (x)\J{x}^{\rho k} (z-H)^{-1} \J{x}^{-\rho(k+1)}u $, we obtain
\begin{align*}
v_l &= (z-H)^{-1}[ H, J_l (x) \J{x}^{\rho k} ] (z-H)^{-1} \J{x}^{-\rho(k+1)}u + (z-H)^{-1}\gamma (\J{x}/l) u 
 \\ & = -  (z-H)^{-1}  \left( \sum_{m=1}^n  x_m \J{x}^{-\alpha /2 -1} p_m + O(1) + O(l^{-1- \alpha /2})\right) \times B_0
  \J{x}^{\rho k} (z-H)^{-1} \J{x}^{- \rho(k+1)} u \\ & \qquad \quad  + (z-H)^{-1} \gamma (\J{x}/l ) u.
\end{align*}
By Lemma \ref{La1} and the assumption of $  \D{\J{x}^{\rho k}} \subset (z-H)^{-1} \D{\J{x}^{\rho k} } $, we have $u_l$ converges as $l \to \infty$, and this means that  $\D{\J{x}^{\rho (k+1)} } \subset (z-H)^{-1} \D{\J{x}^{\rho (k+1)}}$ holds. Then, the property of domain invariance \eqref{ad4} follows from \eqref{ad3} with $e^{-it \cdot } \varphi (\cdot ) \in C_0^{\infty} ({\bf R})$ for any fixed $t$. 

}

Finally, we obtain the property of domain invariance  of $\D{p^2 + |x|^{\alpha} }$, which plays a fundamental role in the analysis of many terms. 
\begin{prop}\label{Pa3}
Let $\CAL{N}_{\alpha} := p^2 + |x|^{\alpha} $. Then, we have 
\begin{align}\label{2}
(z-H)^{-1} \D{\CAL{N}_{\alpha}} \subset \D{\CAL{N}_{\alpha}}, \quad  \varphi (H) \D{\CAL{N}_{\alpha}} \subset \D{\CAL{N}_{\alpha}} .
\end{align}
\end{prop}
\Proof{
First, we demonstrate that $(z-H)^{-1} \D{\CAL{N}_{\alpha}} \subset \D{|x|^{\alpha} }$. Consider that $N$ in Lemma \ref{La2} is sufficiently large such that $ \rho N = 2 + \delta $ and $0\leq \delta <1$, i.e., $\rho N \geq 2>\alpha $. Then, it follows that 
\begin{align*}
\left\| (z-H)^{-1}  \right\| = \left\| \J{x}^0 (z-H)^{-1} \J{x}^{-0} \right\| \leq C_0, \quad  \left\| \J{x}^{2+\delta} (z-H)^{-1} \J{x}^{-2- \delta} \right\| \leq C_{2+ \delta}. 
\end{align*}
The interpolation theorem (see Kato \cite{Ka}) states that for any $0 < \beta < 1$, 
\begin{align*}
 \left\| \J{x}^{(2+ \delta)\beta} (z-H)^{-1} \J{x}^{-(2 + \delta)\beta} \right\| \leq C_0^{1-\beta}C_{2 + \delta}^{\beta}. 
\end{align*}
Taking $\beta = \alpha/(2 + \delta) \in (0,1) $, we obtain the bound of the operator norm of $\J{x}^{\alpha} (z-H)^{-1} \J{x}^{-\alpha} $. Hence, $(z-H)^{-1} \D{\CAL{N}_{\alpha}} \subset \D{|x|^{\alpha} }$ holds. Moreover, from Lemma \ref{La1}, we note $(z-H)^{-1}\D{\CAL{N}_{\alpha}}\subset \D{p^2}$ because $ (z-H)^{-1}\D{\CAL{N}_{\alpha}}\subset \D{\J{x}^{\alpha} }$. 

}

\begin{rem}
In addition, we can show $(z-H)^{-1}\D{p^2 + |x|^{\theta}} \subset \D{p^2 + |x|^{\theta}} $ for any $\theta \geq \alpha$. 
\end{rem}

\section{Large velocity estimate}
In this section, we present the large-velocity propagation estimate for $H$. This type of estimate has already been shown in Proposition 5.7 of \cite{BCHM} with a compactly supported cut-off. This section aims to extend this result to cut-offs that are not compactly supported. This extended result enables us to demonstrate the a key estimate \eqref{ad30}.

In the following, we set $\CAL{N}_{\alpha}= p^2 + |x|^{\alpha}$, and $\varphi \in C_0^{\infty} ({\bf R})$ satisfies $0 \leq \varphi \leq 1$, $\varphi (s) = 0$ for $|s| \leq R -1$ and $\varphi (s) = 0$ for $ |s| \geq R$, where $R$ is a positive constant provided later. Before considering the large velocity estimate, we note that $H$ has no embedded eigenvalues on ${\bf R}$. Hence, considering the cut-off $\varphi \in C_0^{\infty} ({\bf R}) $ instead of $\varphi \in C_0^{\infty} ({\bf R} \backslash \sigma_{\mathrm{pp}}(H) )$ is sufficient. In the following, we therefore can omit the discussion of issues arising from the embedded eigenvalues.  

The following lemma provides the momentum bound under the energy cut-off.

\begin{lem} \label{L1}
We define $A_{0,R} := \left(\alpha \sqrt{n} (2n+1) + \sqrt{\alpha ^2 n (2n+1)^2 + 4 a_{0,R}} \right)/2$, where $a_{0,R} = n(n+1)(\alpha ^2 + 3 \alpha) +n(R+ C_{V,0} + \sigma) $. Then, for all $\phi \in L^2({\bf R}^n)$, 
\begin{align*}
 \sum_{j=1}^n \left\| 
\J{x}^{-\alpha /2} p_j  \varphi (H) \phi
\right\| ^2 \leq A_{0,R}^2 \left\| \varphi (H) \phi \right\|^2 
\end{align*}
holds. In particular, for all $j \in \{1,2,...,n \}$, 
\begin{align*}
\left\| 
\J{x}^{-\alpha /2} p_j  \varphi (H) \phi
\right\|  \leq A_{0,R} \left\| \varphi (H) \phi \right\| 
\end{align*}
and 
\begin{align*}
\left\| 
\J{x}^{-\alpha /2} p_j  (H + i)^{-1} \phi
\right\|  \leq A_{0,1} \left\| \phi \right\| 
\end{align*}
hold.
\end{lem}
\Proof{
For $\phi \in \D{\CAL{N}_{\alpha}}$, we define 
\begin{align*}
I_j := \left\| 
\J{x}^{-\alpha /2} p_j  \varphi (H) \phi
\right\|.
\end{align*}
Then 
\begin{align*}
I_j ^2 \leq \left\| \varphi (H) \phi \right\| \left\| p_j \J{x}^{-\alpha} p_j \varphi(H) \phi \right\|. 
\end{align*}
is obtained. Let $v = \varphi(H) \phi$ and $a_{0,R} = n(n+1)(\alpha ^2 + 3 \alpha) + n(R + C_{V,0}+ \sigma) $. Using $\| (p^2 - \sigma |x|^{\alpha} ) v\| \leq 
\left( R +C_{V,0} \right)  \| v \|$ and
\begin{align*}
& \sum_{j=1}^n \left\| p_j \J{x}^{-\alpha} p_j v \right\| \\  &= \sum_{j=1}^n \left\| \left( -i \alpha p_j x_j \J{x}^{- \alpha -2 }  +  p_j^2 \J{x}^{-\alpha}  \right) v \right\| \\ 
& \leq \sum_{j=1}^n  \left\| \left( \alpha \J{x}^{- \alpha -2} - \alpha (\alpha +2) x_j^2 \J{x}^{- \alpha -4} \right) v \right\| + n \left\| \sum_{j=1}^n p_j ^2 \J{x}^{- \alpha} v \right\| + \alpha \sum_{j=1}^n \left\| x_j \J{x}^{- \alpha -2} p_j v \right\| \\ 
& \leq n (n+1)(\alpha ^2 + 3 \alpha) \|v \|+ \alpha(2n+1) \sum_{j=1}^n \left\| x_j \J{x}^{- \alpha -2} p_j v \right\| \\ & \qquad 
+ n \left\| \J{x}^{- \alpha} \left( p^2 - \sigma |x|^{\alpha} +V  + \sigma |x|^{\alpha} -V \right) v \right\| \\ 
& \leq a_{0,R} \| v \| + \alpha(2n+1) \sum_{j=1}^n \left\| \J{x}^{- \alpha /2} p_j v \right\|, 
\end{align*}
we have 
\begin{align*}
\sum _{j=1}^n I_j^2 &\leq \alpha (2n+1) \| \varphi (H) \phi \| \sum_{j=1}^n I_j +  a_{0,R} \| \varphi (H) \phi \|^2
\\ & \leq \alpha \sqrt{n} (2n+1) \| \varphi (H) \phi \| \left(  \sum_{j=1}^n I_j ^2 \right)^{1/2}+  a_{0,R}\| \varphi (H) \phi \|^2. 
\end{align*}
By solving this inequality with $ \sum I_j^2 \geq 0$, we obtain the desired result. 
}

In addition, we set $\chi$ as a smooth cut-off such that $0 \leq \chi \leq 1$, and for some positive constant $a>0$, $\chi(t) = 1$ for $|t| \geq 2a$ and $\chi(t) = 0$ for $|t| \leq a$. Then, the following estimate holds. 
\begin{lem}\label{L2}
Let $t \geq 0$. For constant $C >0$, the estimate
\begin{align} \label{1}
\left\| 
|x|^{\rho} \chi (|x|^{\rho}/(t+1)) e^{-itH} \varphi(H) \J{x}^{- \rho}
\right\| \leq C \J{t}
\end{align}
holds. 
\end{lem}
\Proof{
For $\phi \in \CAL{N}_{\alpha}$, we first calculate
\begin{align*}
\frac{d}{dt} 
\left( 
e^{itH} |x|^{\rho} \chi (|x|^{\rho} /(t+1)) e^{-itH} 
\right) \varphi (H) \phi &= 
e^{itH} i [ p^2 , |x|^{\rho} \chi (|x|^{\rho} /(t+1)) ] e^{-itH}  \varphi (H) \phi \\ & \quad - e^{itH} \frac{|x|^{2 \rho} }{(t+1)^2} \chi' (|x|^{\rho} /(t+1))e^{-itH}  \varphi (H) \phi 
\\ &= \sum_{j=1}^n  e^{itH} i [ p^2_j , |x|^{\rho} \chi (|x|^{\rho} /(t+1)) ] e^{-itH}  \varphi (H) \phi
\\ & \quad - e^{itH} \frac{|x|^{2 \rho} }{(t+1)^2} \chi' (|x|^{\rho} /(t+1))e^{-itH}  \varphi (H) \phi.
\end{align*}
From the commutator calculation, we have
\begin{align*}
& i[p_j ^2 , |x|^{\rho} \chi(|x|^{\rho} /(t+1) )] \\  &= -i \left( \rho |x|^{\rho -2} + \rho (\rho -2) x_j^2 |x|^{\rho -4}  \right) \chi(|x|^{\rho} /(t+1) )   \\ & \quad  \quad -i \left( 
\frac{2\rho ^2 x_j^2 |x|^{2 \rho -4}}{t+1} 
 + \frac{2 \rho (\rho -1) x_j^2 |x|^{2 \rho -4} }{t+1} + \frac{\rho |x|^{2 \rho -2} }{(t+1)^2} \right) \chi'(|x|^{\rho}/(t+1)) 
 \\ & \qquad \quad -i \frac{\rho ^2x_j ^2 |x|^{3\rho -2}  }{(t+ 1)^2} \chi ''(|x|^{\rho} /(t+1) )
 \\ & \qquad \qquad + 2\left( 
 \rho x_j |x|^{\rho -2} \chi (|x|^{\rho}/t) + \frac{ \rho x_j |x|^{2\rho -2}}{t+1} \chi '(|x|^{\rho}/(t+ 1))
 \right) p_j \\ 
 &=: J_1 + J_2 + J_3 + J_4.
\end{align*}
Clearly, $J_1$, $J_2$, and $J_3$ are bounded operators. Moreover, from Lemma \ref{L1}, we have 
\begin{align*}
\left\| J_4 \varphi(H) \right\| \leq C \left\| \J{x}^{- \alpha /2} p_j \varphi (H) \right\| 
\end{align*}
using $\rho -1 = - \alpha /2$ and $|x|^{\rho} \leq 2a (t+1) $ on $\chi '$. Consequently, we obtain 
\begin{align*}
& \left\| 
e^{itH} |x|^{\rho} \chi (|x|^{\rho} /(t+1)) e^{-itH} \varphi (H) \phi
\right\| \\ & \leq  \left\| 
e^{i 0 H} |x|^{\rho} \chi (|x|^{\rho}) e^{-i 0 H}  \varphi (H) \phi
\right\| + \int_0^t \sum_{j=1}^n  \left\| 
e^{itH} i [ p^2_j , |x|^{\rho} \chi (|x|^{\rho} /(t+1)) ] e^{-itH}  \varphi (H) \phi 
\right\| dt \\ 
& \leq C \| \J{x}^{\rho} \varphi(H) \phi \| + C t \| \phi \|
\\ & \leq C \J{t} \| \J{x}^{\rho} \phi \|,  
\end{align*}
using Lemma \ref{La1} and the Hellfer-Sj\"{o}strand formula.

}

\subsection{Large-velocity estimate}
Let $A_{1,R} = 4 n \rho A_{0,2R}$. Here, we set $F (\cdot) $ as a smooth cut-off such that $0 \leq F \leq 1$, $F(s) = 0$ for $ s \leq A_{1,R} $ and $F(s) = 1$ for $s \geq 2A_{1,R}$. Moreover, we set $G(s) = \int_{-\infty}^s F(\tau) ^2 d \tau$. The purpose of this subsection is to show the following large-velocity estimate. 
\begin{prop}\label{P2}
The inequality 
\begin{align*}
\int_1^{\infty} \left\| 
F\left( \frac{|x|^{\rho}}{t} \right) e^{-itH} \varphi (H)  \J{x}^{- \rho} \phi
\right\|^2 \frac{dt}{t} \leq C \| \phi \|^2
\end{align*}
holds for all $\phi \in L^2({\bf R}^n)$.
\end{prop}
\Proof{
First, we set an observable as
\begin{align*}
\Phi (t) := \J{x}^{- \rho}\varphi(H) e^{itH} G(|x|^{\rho}/t) e^{-it H} \varphi(H) \J{x}^{- \rho}.
\end{align*}
Here, we note that by the definition of $G$, we have 
\begin{align*}
G(s) = 
\begin{cases}
0, & s \leq A_{1,R}, \\ 
\mathrm{(bdd)}, & A_{1,R} < s < 2 A_{1,R}, \\ 
(|s| -2A_{1,R}) + \mathrm{(bdd)}, & s \geq 2 A_{1,R}, 
\end{cases}
\end{align*}
where $\mathrm{(bdd)}$ indicates a bounded function whose norm is bounded by a constant independent of $s$. Hence, we can write $\Phi (t)$ as 
\begin{align*}
\Phi (t) \leq (\mathrm{bdd}) + \J{x}^{- \rho}\varphi(H) e^{itH} \frac{|x|^{\rho}}{t} {\chi} (|x|^{\rho}/(t+1)) e^{-it H} \varphi(H) \J{x}^{- \rho} 
\end{align*}
with a suitable cut-off ${\chi}$ in Lemma \ref{L2}, and hence, by Lemma \ref{L2}, we can determine that $\Phi (t)$ is bounded and whose bound is independent of $t$. 

We now prove this proposition. Straightforward calculations show that: 
\begin{align*}
{\bf D}_{H} (G(|x|^{\rho}/t)) &= \frac{d}{dt} G(|x|^{\rho}/t) + i [H , G(|x|^{\rho}/t)]
\\ &= 
- \frac{|x|^{\rho} }{t^2} F\left( \frac{|x|^{\rho}}{t} \right)^2 + \sum_{j=1}^n \left( \rho \frac{x_j }{t |x|^{2-\rho} } p_j  F\left( \frac{|x|^{\rho}}{t} \right)^2  + \rho F\left( \frac{|x|^{\rho}}{t} \right)^2   p_j \frac{x_j }{t |x|^{2-\rho} }   \right)
\\ & \leq - \sum_{j=1}^n \frac{1}{t} F\left( \frac{|x|^{\rho}}{t} \right) \left(  A_{1,R}  - 2 \rho \frac{x_j}{|x|^{2- \rho}} p_j  \right)F\left( \frac{|x|^{\rho}}{t} \right) + t^{-2} B_0. 
\end{align*}
Moreover, by setting ${\varphi}_0 \in C_0^{\infty} ({\bf R})$ such that $\varphi = {\varphi}_0 \varphi$ and $|t| \leq 2R$ on the support of ${\varphi}_0(t)$, we determine that, for $v(t) = e^{-itH} \varphi(H) \J{x}^{- \rho} \phi$,
\begin{align*}
\left\| 
\J{x}^{- \alpha /2} p_j   F\left( \frac{|x|^{\rho}}{t} \right) v(t)
\right\| &\leq \left\| 
\J{x}^{- \alpha /2} p_j   F\left( \frac{|x|^{\rho}}{t} \right) {\varphi}_0 (H)  v(t)
\right\| 
\\ & \leq C|t|^{-1}  \left\| v(t)
\right\|  + \left\| 
\J{x}^{- \alpha /2} p_j    {\varphi}_0 (H) \right\| \left\| F\left( \frac{|x|^{\rho}}{t} \right) v(t)
\right\| 
\\ & \leq C|t|^{-1}  \left\| v(t)
\right\|  + A_{0,2R} \left\| F\left( \frac{|x|^{\rho}}{t} \right) v(t)
\right\|,
\end{align*}
using
\begin{align*}
\left\| \J{x}^{- \alpha /2} p_j [F(|x|^{\rho}/t) , {\varphi}_0 (H) )] \right\| = \CAL{O}(t^{-1}). 
\end{align*}
Indeed, using the Helffer-Sj\"{o}strand formula, for $\phi \in \D{\CAL{N}_{\alpha}}$, we have 
\begin{align*}
& \J{x}^{- \alpha /2} p_j [F(|x|^{\rho}/t) , {\varphi}_0 (H) ] \phi  \\ & = - \frac{1}{2 \pi i} \int_{\bf C} ( \overline{\partial _z }\tilde{\varphi}_0 (z))  \J{x}^{- \alpha /2} p_j  (z- H)^{-1}[H,F(|x|^{\rho}/t) ] (z- H)^{-1} \phi   dz d \bar{z} \\ &= 
- \frac{1}{2 \pi i} \int_{\bf C} ( \overline{\partial _z }\tilde{\varphi}_0 (z))  \J{x}^{- \alpha /2} p_j  (z- H)^{-1}
\\ & \qquad \qquad \times \left( \left( \frac{B_0}{t^{2/ \rho}} + \frac{2 \rho x_j |x|^{\rho -2}  }{t} \right) F'(|x|^{\rho}/t) p_j \right)  (z- H)^{-1} \phi   dz d \bar{z} 
\\ & = t^{-1} B_0, 
\end{align*}
where $\tilde{\varphi}_0(z)$ denotes the almost analytic extension of $\varphi_0$ (see Helffer-Sj\"{o}strand \cite{HS}). 
Consequently, we obtain 
\begin{align*}
& \frac{d}{dt} \left( \Phi(t) \phi, \phi \right) \\ & \leq - \frac{A_{1,R}}{t} \left\| F\left( \frac{|x|^{\rho}}{t} \right) v(t) \right\|^2 + \frac{2 \rho}{t} \sum_{j=1}^n \left\| F\left( \frac{|x|^{\rho}}{t} \right) v(t)\right\|\left\| \J{x}^{- \alpha /2}p_j F\left( \frac{|x|^{\rho}}{t} \right) v(t) \right\| + \CAL{O}(t^{-2})\| \phi \|^2
\\ & \leq 
- \frac{ A_{1,R} - 2 n \rho A_{0,2R} }{t} \left\| F\left( \frac{|x|^{\rho}}{t} \right) v(t)\right\| ^2 +  C t^{-2}  \| \phi \|^2,
\end{align*}
which yields 
\begin{align*}
2n \rho A_{0,2R} \int_1^{\infty} \left\| F\left( \frac{|x|^{\rho}}{s} \right) v(s) \right\| \frac{ds}{s} &\leq \| \phi \|^2 \int_1^{\infty} \CAL{O}(s^{-2}) ds + \| \Phi(1) \phi \| \| \phi \| + \sup_{t \geq 1 }\| \Phi(t) \phi \| \| \phi \|
\\ & \leq C \| \phi \|^2.
\end{align*}

}

\section{Mourre theory}
In this section, we introduce the Mourre theory for $H$. We set $\CAL{N}_{\alpha} = p^2 + |x|^{\alpha} $. For $\SCR{D}(\CAL{N}_{\alpha})$, we have ${\displaystyle \overline{p^2 - \sigma |x |^{\alpha} } = \overline{p^2} - \overline{\sigma |x|^{\alpha}}}$ because $\SCR{D} (\CAL{N}_{\alpha})= \D{p^2} \cap \D{|x|^{\alpha}}$ holds, and $\D{H} \cap \D{\CAL{N}_{\alpha}} = \D{\CAL{N}_{\alpha}}$ is a core for $H$. Moreover, from Proposition \ref{Pa3}, $\varphi (H_0) \SCR{D}(\CAL{N}_{\alpha}) \subset \D{\CAL{N}_{\alpha}}$ holds. Based on this, the form $i[H, \SCR{A}]$ on $\D{\CAL{N}_{\alpha}}$ has a unique extension to a continuous sesquilinear form on $\D{H}$. We denote this extension by the same notation $i[H, \SCR{A}]$.

 From the compactness argument, we find that for any compact operator $C_0$, there exists $\varphi \in C_0^{\infty} ({\bf R})$ such that
\begin{align*}
\left\| C_0 \varphi (H) \right\| \ll \alpha _0 , \quad \alpha _0 := \mathrm{min} \{  \alpha \sigma , (2- \alpha ) \sigma \} 
\end{align*} 
holds. In the following, we always consider $\| C_0 \varphi (H)\|$ to be sufficiently small compared with other constants.

We first provide a short sketch of the Mourre estimate for the case where $\alpha \geq 1$. The Mourre estimate for $H_0$ was first obtained by \cite{BCHM}, and \cite{It2} then also considered the Mourre estimate using a different conjugate operator to show the Besov bounds for resolvents. In this study, we handle the different types of conjugate operators addressed in \cite{BCHM} and \cite{It2}. In \cite{BCHM}, the authors defined the pseudo-differential operator $\SCR{A}_0$ with symbols, 
\begin{align*}
{a}_{0} (x, \xi) := x \cdot \xi \J{x}^{- \alpha} \psi \left( \frac{\xi^2 - \J{x}^{\alpha}}{\xi ^2 + \J{x}^{\alpha}} \right).
\end{align*}
We note that 
\begin{align*}
\varphi(\tilde{H}_0) i [\tilde{H}_0, \SCR{A}_0 ] \varphi(\tilde{H}_0 ) \geq  \tilde{\delta }  \varphi(\tilde{H}_0 )^2+ C_0,   \quad \tilde{H}_0 := p^2 - \sigma \J{x}^{- \alpha},
\end{align*} 
where $\psi \in C_0^{\infty}({\bf R})$ is narrowly supported around $0$ and $\tilde{\delta} >0$ is a constant; see Lemma 3.16 in \cite{BCHM}. On the other hand, in this study, we set $\SCR{A} := \J{x}^{- \alpha} x \cdot p + p \cdot x \J{x}^{- \alpha} $ and show that  
\begin{align} \label{4}
\varphi({H}_0) i [{H}_0, \SCR{A} ] \varphi({H}_0 ) \geq  (2-  \sigma ) \sigma  \varphi({H}_0 )^2+ C_0, 
\end{align}
where we note that, for $\psi \in \D{\CAL{N}_{\alpha}}$, the commutator $(i[H_0, \SCR{A}] \varphi(H_0) \psi, \varphi(H_0) \psi )$ can be defined in the form sense because $\D{\CAL{N}_{\alpha}} \subset \D{\SCR{A}}$ holds. The calculation in the proof of Proposition \ref{P3} shows that 
\begin{align} \label{3}
\varphi({H}_0) i [{H}_0, \SCR{A}] \varphi({H}_0 ) \geq  \varphi(H_0) \J{x}^{- \alpha /2} \left( 4(1- \alpha) p^2 + 2 \alpha \sigma |x|^{\alpha}  \right) \J{x}^{- \alpha /2}\varphi({H}_0 )+ C_0.
\end{align}
Let $\eta >0$ be a small constant such that $2 - \alpha   -2 \eta > \eta$. Then, the inequality above can be rewritten as 
\begin{align*}
& \varphi({H}_0) i [{H}_0, \SCR{A}] \varphi({H}_0 ) \\ & \geq  \varphi(H_0) \J{x}^{- \alpha /2} \left(  4\eta  p^2 + (4 -2 \alpha  -4 \eta) \sigma |x|^{\alpha} +4(1- \alpha - \eta) H_0 \right) \J{x}^{- \alpha /2}\varphi({H}_0 )+ C_0.
\end{align*}
 If $(4|1- \alpha -  \eta|)H_0 \leq  \eta (p^2 + |x|^{\alpha}) $ holds in the form sense, we can roughly deduce the positivity of the commutator. Hence, \cite{BCHM} employed a cut-off of $\psi (\xi ^2 - \J{x}^{\alpha} /(\xi ^2 + \J{x}^{\alpha}))$. The advatage of using such a cut-off is that we can deduce the positivity of the commutator and that such a commutator can be extended to a bounded operator. Boundedness of commutator enables many easy calculations for deducing the Mourre theory and  propagation estimates. However, such a cut-off makes the commutator calculation difficult. In our approach, we must calculate the commutator of $H_0$ and $\SCR{A}$ at least $5$ times. To simplify this discussion, we introduce another type of conjugate operator.

 Let us suppose that
 $$ 
 \varphi(H_0) \J{x}^{- \alpha /2} H_0 \J{x}^{- \alpha /2}\varphi({H}_0 ) \leq 2R\varphi(H_0) \J{x}^{- \alpha }\varphi({H}_0 ) +C_0
 $$ 
 holds on the support of $\varphi(H_0)$; we can then deduce from \eqref{3} that:
\begin{align*}
\varphi({H}_0) i [{H}_0, \SCR{A} ] \varphi({H}_0 ) & \geq  \varphi(H_0) \J{x}^{- \alpha /2} \left(  4(1- \alpha )H_0 + (4- 2 \alpha) \sigma |x|^{\alpha}  \right) \J{x}^{- \alpha /2}\varphi({H}_0 )+ C_0
\\  & \geq  \varphi(H_0) \J{x}^{- \alpha /2} \left(  (4- 2 \alpha) \sigma |x|^{\alpha} - 8R \right) \J{x}^{- \alpha /2}\varphi({H}_0 )+ C_0. 
\end{align*}
By setting $\chi \in C_0^{\infty} ({\bf R}^n)$ such that $\chi (s) = 0$ on $s \leq R_0$, $R_0 >0$ and noting that $(1- \chi (|x|)) \varphi(H_0)$ is the compact operator (see Lemma 2.3 in \cite{BCHM}), we obtain the following for large $R_0 \gg R$ such that $ (4 - 2\alpha ) \sigma R_0^{\alpha} \J{R_0}^{- \alpha} \geq 32 R \J{R_0}^{- \alpha} $: 
\begin{align}
\nn \varphi({H}_0) i [{H}_0, \SCR{A} ] \varphi({H}_0 ) &\geq  \varphi(H_0) \J{x}^{- \alpha /2} \chi (|x|) \left( (4-2 \alpha) \sigma  |x|^{\alpha} - 8R \right) 
\chi(|x|)\J{x}^{- \alpha /2} \varphi({H}_0 )+ C_0 
\\ & \geq \label{adf1}  \varphi(H_0)  \chi (|x|)  \left( (4-2 \alpha)  \sigma \left( \frac{R_0^2}{1 + R_0 ^2} \right)^{\alpha/2} - 8R \J{R_0}^{- \alpha} \right)   \chi (|x|)  \varphi(H_0) + C_0 
\\ & \geq \nn (2-\alpha) \sigma  \varphi(H_0)     \chi (|x|)^2  \varphi(H_0) + C_0.
\end{align}
Again, using $(1- \chi (|x|)) \varphi(H_0)$ as the compact operator, we obtain \eqref{4} without pseudo-differential cut-offs. This is our scheme for deducing the Mourre theory.

\begin{prop}\label{P3}
Let the conjugate operator $\SCR{A}$ be
\begin{align*}
\SCR{A} := \J{x}^{- \alpha} x \cdot p + p \cdot x \J{x}^{-\alpha}. 
\end{align*}
Then, 
\begin{align} \label{5}
\varphi(H) i[H, \SCR{A}] \varphi(H) \geq \alpha _0 \varphi(H) ^2 + C_0, 
\end{align}
where $\alpha _0 := \mathrm{min} \{  \alpha \sigma , (2- \alpha ) \sigma \}$.
\end{prop}
\Proof{
We divide $\SCR{A} = \SCR{A}_1 + \SCR{A}_1^{\ast} $. Then, 
\begin{align*}
i[H, \SCR{A}_1] = i[p^2 , \SCR{A}_1] - i[ \sigma |x|^{\alpha}, \SCR{A}_1  ] . 
\end{align*}
Straightforward calculations show that: 
\begin{align*}
i[p^2, \SCR{A}_1] = - \alpha  \left(  p \cdot x \J{x}^{- \alpha -2} + \J{x}^{- \alpha -2} x \cdot p \right) x \cdot p  + 2\J{x}^{- \alpha} p^2
\end{align*}
and 
\begin{align*}
- i[ |x|^{\alpha}, \SCR{A}_1] = \alpha |x|^{\alpha} \J{x}^{- \alpha}.
\end{align*}
With
\begin{align*}
\varphi(H) \J{x}^{- \alpha -2} x_j p_j \varphi(H) &= \varphi(H) \J{x}^{- \alpha /2-2} x_j  \cdot \J{x}^{-\alpha /2} p_j \varphi(H)
\end{align*}
as the compact operator, we have
\begin{align*}
\varphi(H) \J{x}^{- \alpha} p^2 \varphi(H) =  \varphi(H) \J{x}^{- \alpha /2}  p^2 \J{x}^{- \alpha /2} \varphi(H) + C_0, 
\end{align*}
using $[p_j , \J{x}^{- \alpha/2}] = - \alpha x_j \J{x}^{- \alpha /2 -2} /2$. Using a similar calculation, we have 
\begin{align*}
& - \alpha \varphi(H)  \left(  p \cdot x \J{x}^{- \alpha -2} + \J{x}^{- \alpha -2} x \cdot p \right) x \cdot p  \varphi(H)
\\ &= 
 - 2\alpha \varphi(H)  \J{x}^{- \alpha /2 } p \cdot x \J{x}^{-1} \cdot \J{x}^{-1} x \cdot p \J{x}^{- \alpha /2 }  \varphi(H) + C_0. 
\end{align*}
For all $\phi \in L^2({\bf R}^n)$, 
\begin{align}\label{adf2}
\left\| 
\J{x}^{-1} x \cdot p \J{x}^{- \alpha /2 }  \varphi(H) \phi
\right\|^2 &\leq   \left\| |p| \J{x}^{- \alpha /2 }  \varphi(H) \phi \right\|^2
\end{align}
holds. Indeed, let $\rho_j = \J{x}^{-1} x_j $, $\psi \in \D{\CAL{N}_{\alpha}}$, and $\psi _j = p_j \psi $. Then, 
\begin{align*}
\left| 
\J{x} ^{-1} x \cdot p \psi
\right|^2 = \left| \sum_{j=1}^n \rho_j \psi _j \right|^2 \leq   \left| \sum_{j=1}^n | \rho_j || \psi _j | \right|^2 \leq 
  \sum_{l=1}^n | \rho_l | ^2  \sum_{j=1}^n | \psi_j | ^2 
\end{align*}
yields 
\begin{align*}
\left\| 
\J{x} ^{-1} x \cdot p \psi
\right\|^2 & \leq  \int_{{\bf R} ^n} \left( \sum_{l=1}^n | \rho_l | ^2   \sum_{j=1}^n | \psi_j | ^2 \right) dx
\\ & \leq \int_{{\bf R} ^n} \left(   \sum_{j=1}^n | \psi_j | ^2 \right) dx
\\ & = \sum_{j=1}^n ( \psi_j, \psi _j ) = \left\| |p| \psi \right\|^2.
\end{align*}
By the density argument, we have \eqref{adf2}. Using \eqref{adf2}, we have
\begin{align*}
& - 2\alpha \varphi(H)  \J{x}^{- \alpha /2 } p \cdot x \J{x}^{-1} \cdot \J{x}^{-1} x \cdot p \J{x}^{- \alpha /2 }  \varphi(H) 
\\ & \geq -2 \alpha  \varphi(H)  \J{x}^{- \alpha /2 }p^2 \J{x}^{- \alpha /2 } \varphi(H) +C_0. 
\end{align*}
The term associated with $\SCR{A}_1^{\ast}$ can be estimated in a similar manner. Consequently, we obtain 
\begin{align*}
\varphi(H) i[H , \SCR{A}] \varphi(H) \geq \varphi(H) \J{x}^{- \alpha /2} \left( 4(1- \alpha) p^2 + 2 \alpha \sigma |x|^{\alpha} \right)\J{x}^{-\alpha /2} \varphi(H) +C_0. 
\end{align*}
For the case in which $\alpha \leq 1$, the following clearly holds: 
\begin{align*}
\varphi(H) i[H , \SCR{A}] \varphi(H) &\geq 2 \alpha \varphi(H) \J{x}^{- \alpha /2} \sigma |x|^{\alpha} \J{x}^{-\alpha /2} \varphi(H) +C_0 
\\ & \geq 2 \sigma  \alpha |R_0|^{\alpha} \J{R_0}^{- \alpha}  \varphi(H) ^2  +C_0 
\\ & \geq \alpha \sigma \varphi(H)^2 + C_0,
\end{align*} 
using the compactness of $(1- \chi (|x|)) \varphi (H)$. Next, we consider the case in which $\alpha >1$. By $p^2 = H + \sigma |x|^{\alpha}$, we have 
\begin{align} \label{6}
\varphi(H) i[H , \SCR{A}] \varphi(H) \geq \varphi(H) \J{x}^{- \alpha /2} \left( 4(1- \alpha) H + ( 4-2 \alpha) \sigma |x|^{\alpha} \right)\J{x}^{-\alpha /2} \varphi(H) +C_0. 
\end{align}
Set $\tilde{\varphi} \in C_0^{\infty} ({\bf R})$ such that $\varphi \tilde{\varphi} = \varphi$ and $\mathrm{supp}\{ \tilde{\varphi} \} \subset \{ s\, | \, |s| \leq 2R \}$. Then, the Helffer-Sj\"{o}strand's formula yields 
\begin{align*}
\varphi(H) \J{x}^{- \alpha /2} H \J{x}^{-\alpha /2} \varphi(H) &= \varphi(H) \tilde{\varphi_k} (H) \J{x}^{- \alpha /2} H  \J{x}^{-\alpha /2} \varphi(H) 
\\ &=  \varphi(H) \J{x}^{- \alpha /2} \tilde{\varphi} (H)  H \J{x}^{-\alpha /2} \varphi(H)  + C_0 
\\ & \leq 
 2R \varphi(H)  \J{x}^{- \alpha } \varphi(H)  + C_0. 
\end{align*}
According to \eqref{6}, this inequality, $4 -2 \alpha >0$, and \eqref{adf1}, we have 
\begin{align*}
\varphi(H) i[H , \SCR{A}] \varphi(H) \geq  (2 -\alpha) \sigma \varphi(H) ^2 + C_0
\end{align*} 
for $\alpha \geq 1$.

}

\section{Strong propagation estimate for $\SCR{A}$.}
This section demonstrates Theorem \ref{T2} using the results of Skibsted \cite{Sk}. Let $\ep >0$ and set $\chi_0 \in C^{\infty}({\bf R})$ with the following properties: 
\begin{align*}
\chi_0 (x) = 
\begin{cases}
1 & x < -2 \ep , \\ 
0 & x >  - \ep , 
\end{cases}
\quad \frac{d}{dx} \chi_0 (x ) \leq 0 , \quad \chi_0 (x) + x \frac{d}{dx} \chi_0 (x) = \tilde{\chi} _0 (x)^2 ,
\end{align*}
where $\tilde{\chi}_0 \geq 0 $ and $\tilde{\chi}_0 \in C^{\infty}({\bf R})$. Moreover, we define 
\begin{align*}
g(x, \tau) = - \chi (x /\tau)  
\end{align*}
for $\tau >0$; see Definition 2.1 in \cite{Sk} with $\alpha \in {\bf N} $ large enough and $\beta =0$. The key operator in \cite{Sk} is $\SCR{A}(\tau)$, which must satisfy Assumption 2.2 in \cite{Sk}. We set $\SCR{A}(\tau) = \SCR{A} - 3 \ep \tau $ and verify that $\SCR{A}(\tau)$ and $H$ satisfy Assumption 2.2 in \cite{Sk}. Here, the important point is that $H$ is bounded from below, which was employed in \cite{Sk} (see Lemma 2.11 in \cite{Sk}) to show the property of domain-invariance \eqref{adadad1}. However, our model of the Hamiltonian $H$ does not have such a condition. Hence, instead of the lower-boundedness of $H$, we provide a different proof (see Lemma \ref{LL1}). Throughout this section, for the two operators $A$ and $B$, we define ${\rm ad}^k_A(B)$ as ${\rm ad}^0_A(B)=B$ and $ {\rm ad}^k_A(B)=  [{\rm ad}^{k-1}_A(B),A]$ for $k\in{\bf N}$.

The necessary conditions we need to show in our model for strong propagation is the followings (see, Assumption 2.2 in \cite{Sk} with $\beta=0$, $n_0 \in {\bf N}$ is large enough and $B= \CAL{N}_{\alpha} $): \\ 
1. With $\mathrm{ad}_{\SCR{A} (\tau) }^0 (H) = H  $ and $1 \leq n \leq n_0$ the form $i^n \mathrm{ad} ^n_{\SCR{A} (\tau) }(H) $ on $\D{H} \cap \D{\CAL{N}_{\alpha}}$ extends to a symmetric operator with domain $\D{H}$. \\ 
2. ${\displaystyle  \sup _{|s| < 1} \|  He^{\SCR{A} (\tau) s} \|  < \infty } $ for any $\psi \in \D{H} $ and $\tau \geq 0$.  \\ 
3. For any $\tau _1 , \tau _2 \geq 0$, $\SCR{A} (\tau _1) - \SCR{A} (\tau _2) $ is bounded, and the derivative $ \displaystyle{d_{\tau} \SCR{A}(\tau) =   \frac{d}{d \tau} \SCR{A} (\tau) }  $ exists in $\SCR{B}(L^2({\bf R}^n))$. 
For $n \leq n_0 -1$ and $\tau \geq t_0$, the form 
\begin{align*}
i^n \mathrm{ad}^n_{\SCR{A}(\tau)} (d _{\tau} \SCR{A} (\tau)) = i[ i^{n-1} \mathrm{ad}_{\SCR{A} (\tau)}^{n-1} (d _{\tau} \SCR{A} (\tau)) , \SCR{A} (\tau) ] 
\end{align*} 
on $\D{\CAL{N}_{\alpha}}$ extends to a bounded selfadjoint operator on $L^2({\bf R}^n)$.
\\ 
4. For $n \leq n_0$, $\mathrm{ad}^n_{\SCR{A} (\tau)} (H)  (H-i)^{-1} $ and $\mathrm{ad}_{\SCR{A} (\tau) }^n (d_{\tau} \SCR{A} (\tau)) $ are continuous $\SCR{B}(L^2({\bf R}^n ))$-valued function of $\tau \geq 0$. 
\\ 
5. ~~(a) $\mathrm{ad}^{n_0 -1} _{\SCR{A} (\tau)} (d_{\tau} \SCR{A} (\tau))  = \CAL{O}(\tau ^{\kappa _0}) $ as $\tau \to \infty$. \\ 
~~~~~~(b) For $n \leq n_0$, $\mathrm{ad}^{n -1} _{\SCR{A} (\tau)} (d_{\tau} \SCR{A} (\tau))(H-i)^{-1}  = \CAL{O}(1) $ as $\tau \to \infty$  \\ 
~~~~~~(c) For $n \leq n_0$, $\mathrm{ad}^{n } _{\SCR{A} (\tau)} ( H)(H-i)^{-1}  = \CAL{O}(1) $ as $\tau \to \infty$ \\ 
6. For large $\alpha >0$, define $\zeta (t) := (-\SCR{A}(\tau) )^{\frac{\alpha -1}2} \tilde{\chi _0} (\SCR{A} (\tau) / \tau )e^{-itH} \varphi(H) \phi $ for $\phi \in L^2({\bf R}^n)$. Then there exists $C>0$ such that  for any $\phi \in L^2({\bf R}^n)$, 
\begin{align} \label{K-5/21-1}
\int_0^{\infty} \left( {\bf D}_H (\SCR{A} (t) )  \tilde{\varphi}(H) \zeta (t) ,  \tilde{ \varphi }(H) \zeta (t)  \right) dt \leq C \| \phi  \| ^2 .
\end{align}
We check all conditions are fulfilled herein:

\underline{\bf For 1:} First, we demonstrate that, for any $n_0 \in {\bf N}\cup \{0 \}$, $i^{n_0} \mathrm{ad}^{n_0}_{\SCR{A}(\tau)} (H) = i^{n_0} \mathrm{ad}^{n_0}_{\SCR{A}} (H) $ can be extended to the symmetric operator on $\D{H}$. Evidently, $\mathrm{ad}^0_{\SCR{A}(\tau)} (H) = H$ satisfies this assumption. From the previous calculation, (see Proposition \ref{P3}), we find that
\begin{align*}
 {-i} \mathrm{ad}_{\SCR{A}(\tau)}^1(H)  &= - \alpha  \left(  p \cdot x \J{x}^{- \alpha -2} + \J{x}^{- \alpha -2} x \cdot p \right) x \cdot p  + 2\J{x}^{- \alpha} p^2
 + \alpha  \sigma|x|^{\alpha} \J{x}^{- \alpha} \\ & \quad - \J{x}^{- \alpha} (x \cdot \nabla V(x)) + (\mathrm{h.c.})
\end{align*}
can be extended to the symmetric operator on $\D{H}$, using the notation $A + A^{\ast} = A + (\mathrm{h.c.})$ for operator $A$. Continuing similar calculations, we can obtain that $\mathrm{ad}^{n_0}_{\SCR{A}(\tau)} (H)$ can be extended to the symmetric operator on $\D{H}$. 

\underline{\bf For 2:} Next, we demonstrate that $\| H e^{-is \SCR{A}} (H +i)^{-1} \| \leq C$ for any $s \in [0,1]$. Let $\psi \in \SCR{S}({\bf R}^n)$ with $\SCR{F}[\psi] \in C_0^{\infty} ({\bf R}^n)$. Then, there exists $\Psi \in C_0^{\infty} ({\bf R})$ such that $\psi = \Psi (p^2) \psi $.  The pseudo-differential operator can then be defined as follows:
\begin{align*}
\CAL{A}(s) \psi := \int_{{\bf R}^n} e^{ix \cdot \xi} a(s; x,\xi) \hat{\psi}  (\xi) d \xi , \quad a(s; x, \xi) = e^{-i s (\J{x}^{- \alpha} x \cdot \xi + \xi \cdot x \J{x}^{- \alpha})} \Psi (\xi).
\end{align*}
Then, noting that $ \| |x|^{\alpha}  \J{x}^{-2} \| \leq C $, the bound 
$$ \| H e^{-i s \SCR{A} } \psi \| \leq C \| (1+p^2+x^2) \psi \| < \infty  $$ 
can be obtained. 
Let $u,v \in \SCR{S} ({\bf R}^n)$ with  $\SCR{F}[u],\SCR{F}[v] \in C_0^{\infty} ({\bf R}^n)$ to consider the form
\begin{align}\label{adadad2}
\left( H e^{- is \SCR{A}}u,  e^{- is \SCR{A}}v   \right) = \left( e^{ is \SCR{A}}H e^{- is \SCR{A}}u,  v   \right). 
\end{align}
By the above argument, we note that $(\CAL{N}_2 -i)^{-1}e^{ is \SCR{A}}H e^{- is \SCR{A}} (\CAL{N}_2 +i)^{-1} $ is strongly differentiable in $s$ and that its derivative is integrable over $[0,s]$. We determine that 
\begin{align*}
&(\CAL{N}_2 -i)^{-1}e^{ is \SCR{A}}H e^{- is \SCR{A}} (\CAL{N}_2 +i)^{-1} 
\\ & = (\CAL{N}_2 -i)^{-1}H (\CAL{N}_2 +i)^{-1} - \int_0^s (\CAL{N}_2 -i)^{-1}e^{ i\tau  \SCR{A}} i[H, \SCR{A}] e^{- i\tau \SCR{A}} (\CAL{N}_2 +i)^{-1} d \tau.
\end{align*}
Then, the equation \eqref{adadad2} is equivalent to  
\begin{align*}
\left( \left( e^{ is \SCR{A}}H e^{- is \SCR{A}}u -Hu + \int_0^s e^{ i\tau  \SCR{A}} i[H, \SCR{A}] e^{- i\tau \SCR{A}} u d \tau \right),  v   \right) =0. 
\end{align*}
Because $v$ can be taken arbitrarily in $\D{\CAL{N}_2}$, for all $\psi \in \SCR{S}({\bf R}^n)$ with $\SCR{F}[\psi] \in C_0^{\infty} ({\bf R}^n)$, we have 
\begin{align*}
\left\| H e^{-i s \SCR{A}} \psi \right\| &= \left\| e^{i s \SCR{A}} H e^{-i s \SCR{A}} \psi \right\|
\\ & \leq \left\| H \psi \right\| + \int_0^{s} \left\| e^{i \tau \SCR{A}} i[\SCR{A} , H ] e^{-i \tau \SCR{A}} \psi \right\| d \tau
\\ & \leq C \left\| H \psi \right\| + \int_0^{s} \left\| i[\SCR{A} , H ] (H+i)^{-1} \right\|  \left\| (H+i)  e^{-i \tau \SCR{A}} \psi \right\| d \tau
\\ & \leq C \left\| (1 +H) \psi \right\| + C\int_0^{s} \left\|H  e^{-i \tau \SCR{A}} \psi \right\| d \tau.
\end{align*}
The Gronwall inequality shows that 
\begin{align*}
\left\| H e^{-i s \SCR{A}} \psi \right\| \leq C \| (1 +H) \psi \|.
\end{align*}
Because $e^{i s \SCR{A}} H  e^{-i s \SCR{A}}$ is the closed operator and $\SCR{F}^{-1} C_0^{\infty} ({\bf R}^n)$ is dense on $\D{H}$, we have, for all $\phi \in \D{H}$, $\left\| H e^{-i s \SCR{A}} \phi \right\| \leq C \| (1 +H) \phi \|$. 

\underline{\bf For 3--5:} By $\SCR{A} (\tau) = \SCR{A} - 3 \ep \tau$, $d_{\tau} \SCR{A} (\tau) = -3 \ep$ and $\mathrm{ad}_{\SCR{A}(\tau)} (\cdot ) = \mathrm{ad}_{\SCR{A}} (\cdot )  $, all conditions 3--5 can be fulfilled. 

\underline{\bf For 6:} By squeezing the support of $\tilde{\varphi}$, one can find that there exists $\delta _0 >0$ such that 
\begin{align*}
\tilde{\varphi} (H)  {\bf D}_H (\SCR{A} (t) ) \tilde{\varphi} (H) = \tilde{\varphi} (H) \left( i[H, \SCR{A}] - 3 \ep   \right)\tilde{\varphi} (H) & \geq (\alpha _0 - 3 \ep)\tilde{\varphi} (H)^2 + \tilde{\varphi} (H)C_0 \tilde{\varphi} (H)
\\ & \geq \delta_0 \tilde{\varphi} (H)^2 
\end{align*} 
and hence Corollary 2.6. in \cite{Sk} with $B_1(\tau) = 0$ and $B_2(\tau) =\delta_0 \tilde{\varphi} (H)^2 $ proves \eqref{K-5/21-1}.

Finally, we show the property of domain invariance. 
\begin{lem}\label{LL1}
Let $\varphi \in C_0^{\infty} ({\bf R})$. Then, for any $t \in {\bf R}$ and $N \in {\bf N}$, 
\begin{align} \label{adadad3}
e^{-itH} \varphi(H) \D{\SCR{A} ^N} \subset \D{\SCR{A}^N}
\end{align}
holds. Moreover, for all $\psi \in \D{\SCR{A}^N }$, there exist $C_N>0$ such that 
\begin{align}\label{adadad4}
\left\| \SCR{A}^N e^{-itH} \varphi(H)  \psi \right\| \leq C_N t^{N+1} \| \SCR{A}^N \psi  \|.
\end{align}
\end{lem}
\Proof{
Straightforward calculations show that:
\begin{align*}
 & e^{-itH}\varphi(H) (\SCR{A}+ i )^{-N} 
 \\ & =  (\SCR{A} +i)^{-1} \left[ \SCR{A},  e^{-itH}\varphi(H) \right] (\SCR{A} +i)^{-N} +  (\SCR{A} +i)^{-1} e^{-itH}\varphi(H)(\SCR{A} +i)^{1-N}
 \\  & \quad \vdots 
 \\ & = (\SCR{A} + i)^{- N }  \sum_{l_1 =0}^{N} C_{l_1}  \mathrm{ad}_{\SCR{A}}^{N-l_1} (e^{-itH}\varphi(H))  (\SCR{A} + i)^{- N + l_1 }.
\end{align*}
We know that the commutator $ \mathrm{ad}_{\SCR{A}}^{N-l_1} (e^{-itH}\varphi(H)) $ can be defined inductively using Helffer-Sj\"{o}strad's formula; because $ \CAL{C}_{t} (x) := \cos (t x) \varphi( x ) \in C_0^{\infty} ({\bf R})$ (as well as $\CAL{S}_t (x) := \sin (t x) \varphi( x )$), for any fixed $t$, we can apply the Helffer-Sj\"{o}strand formula to $\CAL{C}_t(H)$ and obtain 
\begin{align*} 
\CAL{C}_t (H) = c' \int_{{\bf C}} \overline{\partial _z} \widetilde{c_t} (z) (z-H)^{-1} dz d \bar{z}, 
\end{align*}
where $c' = (2 \pi i )^{-1}$, $\widetilde{c_t} $ is the almost analytic extension of $\CAL{C}_t$, and $\widetilde{c_t } (z)$ is written as 
\begin{align*} 
\widetilde{c_t} (z) = \sum_{k=0}^{N-1} c'_k  \left( \frac{d^k}{dx^k} \left( \CAL{C}_t^{} (x) \right) \right) y^{k} \psi ( y/\J{x} ), \quad z = x +iy, \quad c_k '= \frac{i^k}{k!}.
\end{align*}
Because $\CAL{C}_t \in C_0^{\infty} ({\bf R})$, for any $s >0$ and $N_0 \in {\bf N}$, the following well-known estimate holds:
\begin{align*}
\left|  \overline{\partial _z}  \widetilde{{c}_t} (z)  \right| \leq C t^k |\mathrm{Im}z|^k \J{z}^{-k-s-1}, \quad 1 \leq k \leq N_0-1,
\end{align*}
where $C$ is independent of $t$. Then, it immediately follows that
\begin{align*}
 \left[ \SCR{A}, \cos (tH)\varphi(H) \right]
  &=  c' \int_{{\bf C}} \overline{\partial _z} \widetilde{c_t} (z) [\SCR{A} , (z-H)^{-1} ] dz d \bar{z} 
  \\ &  = c' \int_{{\bf C}} \overline{\partial _z} \widetilde{c_t} (z) (z-H)^{-1} [\SCR{A} , H ] 
  (z- H)^{-1} dz d \bar{z} 
\end{align*}
is a bounded operator (and $\left[ \SCR{A}, \sin (tH)\varphi(H) \right]$ is bounded operator), and it also follows that
\begin{align*}
\left\| 
\left[ \SCR{A}, \cos (tH)\varphi(H) \right]
\right\| \leq C \int_{\bf C} \J{z}^{-s-3} t^2 |\mathrm{Im} z|^2 \| (z-H)^{-1} \|^2 dz d \bar{z} \leq C t^{2}.
\end{align*}
Inductively, we determine that $\mathrm{ad}_{\SCR{A}}^N (\cos (tH) \varphi(H)) $ is defined as the bounded operator which satisfies $\| \mathrm{ad}_{\SCR{A}}^N (\cos (tH) \varphi(H)) \| \leq Ct^{N+1}$ (as well as $\mathrm{ad}^N_{\SCR{A}} (\sin (tH) \varphi(H)) $). This proves \eqref{adadad3} and \eqref{adadad4}.

}
\begin{rem}
The growth order in $t$ in \eqref{adadad4} is much stronger than the result in \cite{Sk} (in \cite{Sk}, the growth order is $t^N$).
\end{rem}

Owing to Corollaries 2.5 and 2.6 in \cite{Sk} and the Mourre inequality, by taking $n_0$ to be sufficiently large, we obtain the following for large $\alpha _0 ' $ and $\psi \in \D{\SCR{A}^{\alpha _0 ' /2} }$: 
\begin{align*}
\left\| 
\sqrt{ \chi_0 ( \SCR{A}(\tau) / \tau  ) } e^{-i \tau H} \varphi  (H ) \psi 
\right\| \leq C \tau ^{ - \alpha _0' /2} \left\| \J{\SCR{A}}^{ \alpha _0' /2} \psi  \right\|.
\end{align*}
By taking $\ep$ as $\delta$, $\tau$ as $t$ and $\alpha_0'$ as $2 \kappa$, we find 
\begin{align*}
\left\|  
\sqrt{ \chi_0 \left( \frac{\SCR{A} - 3\delta t}{t}  \right) } e^{-i t H} \varphi  (H ) \psi 
\right\|  \leq C t ^{ - \kappa } \left\| \J{\SCR{A}}^{ \kappa} \psi  \right\|, 
\end{align*} 
and by taking $\sqrt{\chi_0 ( \cdot -3 \delta ) }$ as $g(\cdot)$, Theorem \ref{T2} can be shown.

\section{Nonexistence of wave operators}
We now prove the main theorem. The proof is divided into two parts: \\ ~~ \\ 
{\bf Case 1:} We prove the case where $\theta = \rho$. \\ 
{\bf Case 2:} We prove the case where $\theta < \rho$. \\ ~~ \\ 
The key argument involves deducing the decay estimate, 
\begin{align*}
\left\|  V e^{-itH_0} \phi \right\| \leq C |t|^{-1},
\end{align*}
for $ \phi \in \SCR{S} ({\bf R}^n)$, and to show this, we employ Theorem \ref{T2}, which proves 
\begin{align*}
\left\|  V e^{-itH_0} \phi \right\| & \leq \left\|  V g(\SCR{A} /t) e^{-itH_0} \phi \right\| +  \left\|  V\left( 1-  g(\SCR{A} /t) \right) e^{-itH_0} \phi \right\|. \\ 
 & \leq C |t|^{-1} + \left\|  V\left( 1-  g(\SCR{A} /t) \right) e^{-itH_0} \phi \right\|. 
\end{align*}
In addition, showing the decay estimate is necessary: 
\begin{align*}
 \left\|  V\left( 1-  g(\SCR{A} /t) \right) e^{-itH_0} \phi \right\| \leq C |t|^{-1} .
\end{align*}
To justify this estimate, $V$ must be decayed as $|V(x)| \leq C \J{x}^{- \rho} $ because $\SCR{A} \J{x}^{- \theta} \varphi (H_0)  $ is an unbounded operator if $\theta < \rho$. 
Hence, we first show Theorem \ref{T1} with $\theta = \rho$. Then, by employing a different approach, we show Theorem \ref{T1} with $\theta < \rho$.
\\ ~~ \\ 
{\bf Proof of Case 1:} \\  We assume that 
\begin{align*}
W^+ := \mathrm{s-} \lim_{t \to \infty} e^{itH} e^{-itH_0}
\end{align*}
exists and that it leads to a contradiction. Let $t_2 >t_1 \gg 1$ and
\begin{align*}
Y(t_1, t_2) := \left( 
\left( 
e^{it_2H} e^{-it_2 H_0} - e^{it_1H} e^{-it_1 H_0}
\right) \phi, W^+ \phi
\right), 
\end{align*}
where we set $\phi \in \SCR{S}({\bf R}^n)$ such that $\phi = \varphi (H_0) \phi$ with $\varphi$ defined as in \S{2}. Then, $Y(t_1,t_2)$ can be estimated as follows: 
\begin{align*}
|Y(t_1, t_2) | &= \left| 
\int_{t_1}^{t_2} \frac{d}{dt} \left( 
e^{itH} e^{-itH_0} \phi, W^+ \phi
\right) dt
\right| 
\\ &= 
\left| 
\int_{t_1}^{t_2} \left( 
e^{itH} V e^{-itH_0} \phi, W^+ \phi
\right) dt
\right| 
\\ & \geq |J_1| -|J_2|
\end{align*}
with 
\begin{align*}
J_1 = 
\int_{t_1}^{t_2} \left( V e^{-itH_0} \phi , e^{-itH_0} \phi \right) dt
\end{align*}
and
\begin{align*}
J_2 = \int_{t_1}^{t_2} \left( 
V e^{-itH_0} \phi , e^{-itH} \left( W^+ - e^{itH} e^{-itH_0} \right) \phi
\right) dt. 
\end{align*} 
By Assumption \ref{A1} on $V$, for all $\psi \in L^2({\bf R}^n)$, we have 
\begin{align*}
| (V \psi, \psi ) | \geq c_0 \left| \left( \J{x}^{- \rho} \psi, \psi \right) \right|, 
\end{align*}
which yields the following for $\tilde{F}(s) = {1 - F (s) }$ with the $F$ in Proposition \ref{P2}: 
\begin{align*}
c_0 ^{-1}|J_1| &\geq 
\int_{t_1}^{t_2} \left( \J{x}^{- \rho} \tilde{F}(|x|^{\rho}/t) e^{-itH_0} \phi ,  
 \tilde{F}(|x|^{\rho}/t) e^{-itH_0}
\right) dt.
\end{align*}
On $\tilde{F}(|x|^{\rho}/t)$, $|x|^{\rho} \leq 2A_{1,R} t$ holds. Therefore, we have 
\begin{align*}
|J_1| &\geq c_0 (2A_{1,R} + 1)^{-\rho} 
\int_{t_1}^{t_2} \left\| \tilde{F}(|x|^{\rho}/t) e^{-itH_0} \phi \right\| ^2 \frac{dt}{t}. 
\end{align*}
From Proposition \ref{P2} and $\tilde{F} = 1- F$, we have 
\begin{align*}
|J_1| \geq \frac{3c_0 (2A_{1,R} + 1)^{- \rho}}{4} \| \phi \|^2 \int_{t_1}^{t_2} \frac{dt}{t} - C \| \J{x}^{\rho} \phi \|^2, 
\end{align*}
using $|a+b|^2 \geq 3 a^2 /4 -3b^2$. 

Next, we estimate $J_2$. Because we assume that $W^{+}$ exists, for any $\ep_0 >0$ and a sufficiently small constant compared with $A_{1,R}$ and $c_0$, there exists $t_1 >0$ such that for all $t > t_1$, 
\begin{align*}
\left\| e^{-itH} \left( W^{\pm} -e^{itH}e^{-itH_0}  \right) \phi \right\| \leq \ep_0 \| \phi \|. 
\end{align*}
Hence, we have 
\begin{align*}
|J_2| \leq \ep_0 \| \phi \| \int_{t_1}^{t_2} \left\| V( g(\SCR{A}/t ) + (1- g(\SCR{A}/t))) e^{-itH_0} \phi \right\| dt,
\end{align*}
with $g$ as in Theorem \ref{T2}. By Theorem \ref{T2} with $\kappa =5$, the term associated with $g(\SCR{A}/t )$ can be estimated as 
\begin{align} \label{8}
C \ep_0 \| \phi \| \| \J{\SCR{A}}^2 \phi \| \int_{t_1}^{t_2} \frac{dt}{t^2} \leq C \ep_0 \| \phi \| \| \J{\SCR{A}}^{2} \phi \|.
\end{align}
Next, we estimate the term associated with $ (1- g(\SCR{A}/t)))$. We first show the following inequality: 
\begin{align} \label{7}
\| V (1- g(\SCR{A}/t))) \varphi (H_0) \| \leq C t^{-1} .
\end{align}
Using the Helffer-Sj\"{o}strand's formula, boundedness of an operator $[\J{x}^{- \rho} , \SCR{A} ]$, and commutator expansion (see, \S{C.3} in Derezi\'{n}ski-G\'{e}rard \cite{DG}), we have
\begin{align*}
\J{x}^{- \rho} (1 - g(\SCR{A}/t )) \varphi (H_0) = t^{-1} B_0 + (1-g(\SCR{A}/t)) \J{x}^{- \rho} \varphi (H_0).
\end{align*}
On the support of $(1-g(\SCR{A}/t)) $ we obtain 
\begin{align*}
\left\| 
 (1-g(\SCR{A}/t)) V \varphi (H_0)
\right\| \leq \frac{1}{\delta t} \left\| \SCR{A} \J{x}^{- \rho} \varphi (H_0) \right\| \leq C t^{-1} + Ct^{-1}\sum_{j=1}^n \left\| \J{x}^{- \alpha /2} p_j \varphi(H_0)  \right\|.
\end{align*}
Hence, \eqref{7} is obtained. Then, we have 
\begin{align*}
\int_{t_1}^{t_2} \left\| V (1- g(\SCR{A}/t))) e^{-itH_0} \phi \right\| dt \leq C \| \phi \| \int_{t_1}^{t_2} \frac{dt}{t}, 
\end{align*}
which with \eqref{8} yields
\begin{align} \label{9}
\left| J_2 \right| \leq C \ep_0 \| \phi \| \left( \| \phi \| + \left\| \SCR{A}^2 \phi \right\| \right) + C \ep_0 \| \phi \|^2 \int_{t_1}^{t_2} \frac{dt}{t}. 
\end{align}
~~ \\ 
{\bf Conclusion} \\ 
Suppose that $\CAL{W}^{+}$ exists and $\phi \in \SCR{S}({\bf R}^n)$ for $\phi = \varphi (H_0) \phi$. Let $t_1$ be sufficiently large, such that $\ep_0$ in \eqref{9} becomes sufficiently small compared with $3c_0(2A_1 + 1)^{- \rho}$. In these situations, we have, on the one hand: 
\begin{align} \label{10}
\left|
Y(t_1, t_2)
\right| \leq 2 \| \phi \|^2 \leq 2 \| \phi \| \left( \| \J{x}^{\rho} \phi \| + \| \J{\SCR{A}}^2 \phi \| \right), 
\end{align} 
and on the other hand: 
\begin{align}
\nn \left|
Y(t_1, t_2)
\right| &\geq |J_1| - | J_2| 
\\ & 
\label{11}
\geq \left( \frac{3c_0 (2A_1 +1)^{- \rho}}{4} - C\ep_0  \right) \| \phi \|^2 \int_{t_1}^{t_2} \frac{dt}{t} - C  \| \phi \| \left( \| \J{x}^{\rho} \phi \| + \| \J{\SCR{A} } ^2 \phi  \| \right). 
\end{align}
Take $\| \phi \| = 1$, $\| \J{x}^{\rho} \phi \| + \| \J{\SCR{A} }^2 \phi \| = \tilde{C}$. Then, \eqref{10} and \eqref{11} imply that: 
\begin{align*}
\int_{t_1}^{t_2} \frac{dt}{t} \leq C \tilde{C},
\end{align*}
which fails as $t_1 \to \infty$. This contradiction indicates that $\CAL{W}^{+}$ does not exist.
\\ ~~ \\ 
{\bf Proof of Case 2:} \\  Let $H = H_0 +V$ and $H_{\rho } = H_0 + \J{x}^{- \rho } + V $, where $V$ satisfies Assumption \ref{A1} with $0< \theta < \rho$. We assume that two wave operators 
\begin{align} \label{ad20}
W_{\theta}^+ :=   \mathrm{s-} \lim_{t \to \infty} e^{itH} e^{-itH_0}
\end{align}
and 
\begin{align} \label{ad21}
W_{\theta, \rho}^+ :=   \mathrm{s-} \lim_{t \to \infty} e^{itH_\rho} e^{-itH_0}.
\end{align}
exist and lead contradiction. Here, we note that all arguments in the proof of Case 1 are true for the pair $e^{itH} e^{-itH_{\rho}}$ because 
$$
\frac{d}{dt} e^{itH} e^{-itH_{\rho}} = e^{itH} \J{x}^{- \rho} e^{-itH_{\rho}} ,
$$ 
which implies that $\mathrm{s-} \lim_t e^{itH} e^{-itH_{\rho}} $ does not exist. Here, we note that by the density argument, unitarity of $ e^{itH} e^{-itH_{\rho}}$, and arbitrariness of the choice of $\varphi \in C_0^{\infty} ({\bf R})$, we can demonstrate the nonexistence of wave operators as the following sense: 
\begin{align} \label{M-ad1}
\mbox{`` }  \forall u \in L^2({\bf R}^n) \backslash \{0\},\, \nexists v \in L^2({\bf R}^n) \mbox{ s.t. }{\displaystyle \mathrm{s-} \lim_{t \to \infty} e^{itH} e^{-itH_{\rho}} u = v} \mbox{ "}. 
\end{align}
Then, the identity
\begin{align*}
e^{itH} e^{-itH_0} = e^{itH} e^{-itH_{\rho}} \cdot e^{itH_{\rho}} e^{-itH_0}
\end{align*} 
shows that either $W^+_{\theta}$ or $W^{+}_{\theta, \rho} $ do not exist (or neither $W^+_{\theta}$ nor $W^{+}_{\theta, \rho} $ exist). Indeed, if both $W^+_{\theta}$ and $W^{+}_{\theta, \rho} $ exist, then for all $u \in L^2({\bf R}^n)$, there exist $w_{+, \theta} , w_{+, \theta, \rho} \in L^2({\bf R}^n)$ such that 
\begin{align*}
e^{itH}e^{-itH_0}u - w_{+, \theta} \to 0  \quad \mbox{and} \quad  e^{itH_{\rho}}e^{-itH_0} u - w_{+, \theta, \rho} \to 0 
\end{align*} 
hold. Then, the following also follows:
\begin{align*}
0 &= \lim_{t \to \infty} \left\| e^{itH} e^{-itH_0} u - w_{+, \theta} \right\| 
\\ &=  \lim_{t \to \infty} \left\| e^{itH}e^{-itH_{\rho}} \left( e^{itH_{\rho}}e^{-itH_0} u -w_{+,\theta, \rho} \right) + \left( e^{itH}e^{-itH_{\rho}} w_{+,\theta, \rho}- w_{+, \theta} \right) \right\|, 
\end{align*} 
which yields 
\begin{align*}
\lim_{t \to \infty} \left\| e^{itH}e^{-itH_{\rho}} w_{+,\theta, \rho}- w_{+, \theta}  \right\| \leq \lim_{t \to \infty} \left\| e^{itH_{\rho}}e^{-itH_0} u -w_{+,\theta, \rho} \right\| = 0.
\end{align*}
Hence, $e^{itH}e^{-itH_{\rho}} w_{+,\theta, \rho} \to w_{+, \theta}$, which contradicts \eqref{M-ad1}. Because $V + \J{x}^{- \rho} $ satisfies Assumption \ref{A1} with $\theta < \rho$, the nonexistence for  $W^+_{\theta}$ or $W^{+}_{\theta, \rho} $ indicates the nonexistence of both $W^+_{\theta}$ and $W^{+}_{\theta, \rho} $, which is the desired result.

~~ \\  ~~ \\ 
{\bf Acknowledgement} \\  
A. Ishida is partially supported by JSPS KAKENHI Grant Number JP20K03625 and JP21K03279. M. Kawamoto is partially supported by JSPS KAKENHI Grant Number JP20K14328.

\end{document}